\documentclass[aps,prd,preprint,onecolumn,amsmath,amssymb,floatfix]{revtex4-2}
\usepackage{graphicx}
\usepackage{rotating}
\usepackage{color}
\usepackage{soul}
\usepackage{hyperref}
\usepackage{threeparttable}
\usepackage{natbib} 
\usepackage{longtable}
\usepackage{appendix}
\usepackage{array}   
\usepackage{tabularx}
\usepackage[utf8]{inputenc}

\newcommand{\bfp}{\mbox{\boldmath $p$}}

\def\be{\begin{equation}}
\def\ee{\end{equation}}
\def\ba{\begin{eqnarray}}
\def\ea{\end{eqnarray}}
\def\bl#1\el{\begin{align}#1\end{align}}

\begin{document}
\title{A Systematic Study of Magnetic Fields Impacts on Neutrino Transport in Core-Collapse Supernovae}

\author{Y. Luo$^{1}$}
\email{Corresponding author: yudong.luo@pku.edu.cn}
\author{S. Zha$^{2}$}
\email{zhashuai@ynao.ac.cn}
\author{T. Kajino$^{3,4,5}$}
\email{kajino@buaa.edu.cn}

\affiliation{School of Physics and Kavli Institute for Astronomy and Astrophysics, Peking University, Beijing 100871, China}
\affiliation{International Centre of Supernovae (ICESUN), Yunnan Key Laboratory of Supernova Research, Yunnan Observatories, Chinese Academy of Sciences (CAS), Kunming 650216, People's Republic of China}
\affiliation{School of Physics, Peng Huanwu Collaborative Center for Research and Education, and International Research Center for Big-Bang Cosmology and Element Genesis, Beihang University, Beijing 100191, China}
\affiliation{Graduate School of Science, The University of Tokyo, Tokyo 113-0033, Japan}
\affiliation{National Astronomical Observatory of Japan, Tokyo 181-8588, Japan}

\date{\today}

\begin{abstract}
We quantify the impact of strong magnetic fields (assuming $B=B_0\cdot r_0^3/r^3$ with $B_0\gtrsim 10^{16}$\,G) on the neutrino transport in core-collapse supernovae (CCSNe). Magnetic fields quantize the momenta of electrons and positrons, resulting in an enhanced absorption cross section for low-energy neutrinos and suppressed chemical potentials for $e^\pm$.
We include these changes in the M1 scheme for neutrino transport and perform 1-D CCSNe simulations with \texttt{GR1D}. The increased low-energy cross sections reduce the $\bar{\nu}_e$ mean energy $\langle E_{\bar\nu_e}\rangle$ while elevating the neutrino number luminosities $\mathcal{L_\nu}$ for both ${\nu}_e$ and $\bar{\nu}_e$ due to the lower energy weighted spectra. 
The reduction of chemical potential enhances the $\bar{\nu}_e$ emission while suppressing that of $\nu_e$, thereby driving an increase in the electron fraction behind the stalled shock at $\sim30$--$100$\,km. This further amplifies $\langle E_{\nu_e}\rangle$ through an increased electron density. Consequently, magnetic fields amplify $L_{\nu_e}$ by increasing both $\mathcal{L}_{\nu_e}$ and $\langle E_{\nu_e}\rangle$ whereas for $\bar\nu_e$, the rise in $\mathcal{L}_{\bar\nu_e}$ is offset by a decreased $\langle E_{\bar\nu_e}\rangle$, leading to a minimal change in $L_{\bar\nu_e}$.
A systematic parameter scan of dipole field configurations suggests that, for $r_0 > 30$ km, $\langle E_{\bar{\nu}_e} \rangle$ is significantly suppressed and $L_{\nu_e}$ is enhanced if $B_0 \geq {2.7} \times 10^{16}$\,G.  These magnetic effects become negligible for $B_0$ below $\sim {7.4} \times 10^{15}$\,G.
\end{abstract}

\maketitle

\section{Introduction}
Core-collapse supernovae (CCSNe) are powerful explosions that occur when massive stars run out of fuel. The collapse of their iron cores release{s} a tremendous amount of gravitational energy ($\sim10^{53}$ erg) and neutrinos carry away most of the energy~\citep{Bethe:1990mw}. The commonly accepted delayed-explosion mechanism involves several steps: After forming a homologous core, the density keeps increasing, making the core exceed the nuclear matter density ($ \sim 2.7\times10^{14}\,\rm g\ cm^{-3}$)~\cite{oertel17}. Then, the core bounces back and triggers a shock wave at the core surface. However, this shock quickly turns into an accretion shock due to energy loss from dissociating matter and neutrino emissions, and it stalls at $\sim100$--200\,km. The competition between neutrino heating in the so-called ``gain region" and accretion onto the proto-neutron star (PNS) dictates the subsequent dynamics. {F}or a {successful} explosion, neutrino heating wins the competition and further leads to the shock revival~\cite{Bethe:1990mw,2002RvMP...74.1015W,Janka:2006fh}. Otherwise, the PNS continues contraction and collapses into a black hole if the shock revival fails~\cite{sumiyoshi06,oconnor11}.

The numerical description of such a  “delayed neutrino heating” {seldom} realizes {a} self-consistently explosion in a 1D spherical-symmetric simulation, except for the lowest-mass progenitors~\cite{kitaura06,fischer10,radice17}. The success of the delayed supernova mechanism turned out to be sensitive to many aspects such as {the} mass density structure of the unstable core, the neutrino transport, and the turbulence in the gain region. The consideration of these effects requires realistic multidimensional simulations (see~\citet{Janka:2017vlw} and references therein). 

After the core-collapse, a new-born PNS may contain strong magnetic field{s}~\cite{price,takiwaki09,kiuchi14,nakamura15,kiuchi15,Mosta:2015ucs,ruiz,Raynaud:2020ist,Obergaulinger:2020cqq,Mosta:2014jaa,powell23}. 
For a strong magnetic field, $e^\pm$ energy spectrum is quantized, {and this} could further affect neutrino transport via the modification of neutrino-nucleus cross section. Several previous studies investigated the neutrino-nucleon interactions inside the magnetic field of PNS but without investigating the effects in CCSN dynamical simulations~\cite{Lai:1998sz,Arras:1998mv,Duan:2004nc,Duan:2005fc}. 
Recently,~\citet{2021ApJ...906..128K} implemented the theoretical formulations presented by \citet{Arras:1998mv} within a three-dimensional magnetohydrodynamical (MHD) CCSN simulation, demonstrating that magnetic fields of strength $10^{15-16}$\,G yield no significant observable differences. However, due to the costs of 3D simulations, a detailed characterization of the mechanisms through magnetic field and the precise criteria for safely disregarding these effects {have} not yet been systematically elucidated.

This study explore{s} two primary effects of magnetic fields on neutrino transport: first, the quantization of {the} $e^\pm$ phase space modifies the weak interaction cross section; second, the electron chemical potential $\mu_e$ deviates from its field-free value due to this quantization.
{In particular, we include these effects into 1D simulations and elucidate the feedback arising from microscopic physics. We also explore a range of magnetic field configurations to systematically investigate their influence on neutrino properties.}

This paper is arranged as follows. In Section \ref{nu_trans}, we detail our modifications to the neutrino-nucleon interactions in neutrino transport to incorporate magnetic field effects. In Section \ref{results}, we present the simulation results and discuss the impact on the CCSN neutrino emission. Finally, conclusions are provided in Section \ref{con_dis}.  
 
\section{The neutrino transport inside magnetic field}
\label{nu_trans}
In this work, we use the open source \texttt{GR1D} code to simulate 1-D CCSN explosion~\cite{OConnor:2009iuz,OConnor:2014sgn} with a two-moment neutrino transport scheme using the analytic M1 closure (M1 scheme hereafter)~\cite{Shibata:2011kx,OConnor:2012bsj,OConnor:2014sgn}. The LS180 EOS~\cite{Lattimer:1991nc} is applied in the simulation, and we took a 9.6 M$_\odot$ progenitor model with zero metallicity~\cite{Heger_pri}. The detailed explanation of the \texttt{GR1D} and the M1 scheme refers to \citet{OConnor:2014sgn}.
We use the default neutrino-matter interaction library generated by \texttt{NuLib} \footnote{https://github.com/evanoconnor/NuLib} during the collapse phase. The core bounce is defined as the moment when entropy per baryon at the edge of the inner core reaches 3\,$k_{\rm B}$ per nucleon. We switch on the modification of the neutrino absorption cross section due to magnetic fields at 0.02\,s after the bounce (i.e., $t_{pb}>0.02$\,s). In the M1 scheme, the absorption opacities are given as~\cite{OConnor:2014sgn}
\begin{eqnarray}
     \kappa_{\nu_e n} = \sigma_{\nu_e n}(1-f_{e^-}) (1-f^{eq}_{\nu_e}) X_n\rho N_AW_M, \\
     \kappa_{\bar{\nu}_e p} = \sigma_{\bar{\nu}_e p}(1-f_{e^+}) (1-f^{eq}_{\bar{\nu}_e}) X_p\rho N_A W_{\bar{M}} ,
\end{eqnarray}
where $X_n$ and $X_p$ are the neutron and proton fraction, $W_M$ and ${\bar{M}}$ are the weak magnetism correction~\cite{Vogel:1983hi}, respectively. $\sigma_{\nu_e n}$ and $\sigma_{\bar{\nu}_e p}$ are neutrino-nucleon cross sections. The magnetic field modifies these cross sections via Landau quantization of $e^\pm$ phase space (see Appendix for the detailed derivation), and we use the modified cross section $\sigma_{\nu_e N}(B)$ in Eq.~\ref{sigma_B} in the simulation. {Notice in Eq.~\ref{sigma_B}, $\Theta_\nu$ is the angle between neutrino momentum and the magnetic field, its impact turns out to be less than a factor of three for the case with $B<5\times10^{16}$ G in the Leakage scheme \cite{Luo:2024qmq}. Therefore, we ignore the $\Theta_\nu$ dependence of the cross section and fix $\Theta_\nu=0$ in the simulation.}
$f^{eq}$ is the distribution function of neutrinos in the weak equilibrium. The equilibrium neutrino chemical potential satisfies $\mu_{\nu_e} =\mu_e-(\mu_n-\mu_p)=-\mu_{\bar{\nu}_e}$, and in particular, the electron chemical potential $\mu_e$ is replaced by $\mu_e(B)$ in Eq.~\ref{chemi_po}. To reduce the computational cost, we create a table of $\mu_e(\rho Y_e, T, eB)$ based on Eq.~\ref{chemi_po}, and then obtain $\mu_e$ through the Lagrangian interpolation during the simulation. The neutrino emissivity is calculated from Kirchhoff’s law $\eta_\nu=\kappa B_\nu$, where $B_\nu$ is the neutrino blackbody spectrum.
The neutrino energy spectrum is discretized into 18 logarithmically spaced energy groups, where the lowest energy group centers at 1\,MeV with a width of 2\,MeV and the largest one centers at 280.5\,MeV with a width of 61\,MeV.
\begin{figure}
\centering
\includegraphics[width=1.0\textwidth,clip=true,trim=0cm 0cm 0cm 0cm]{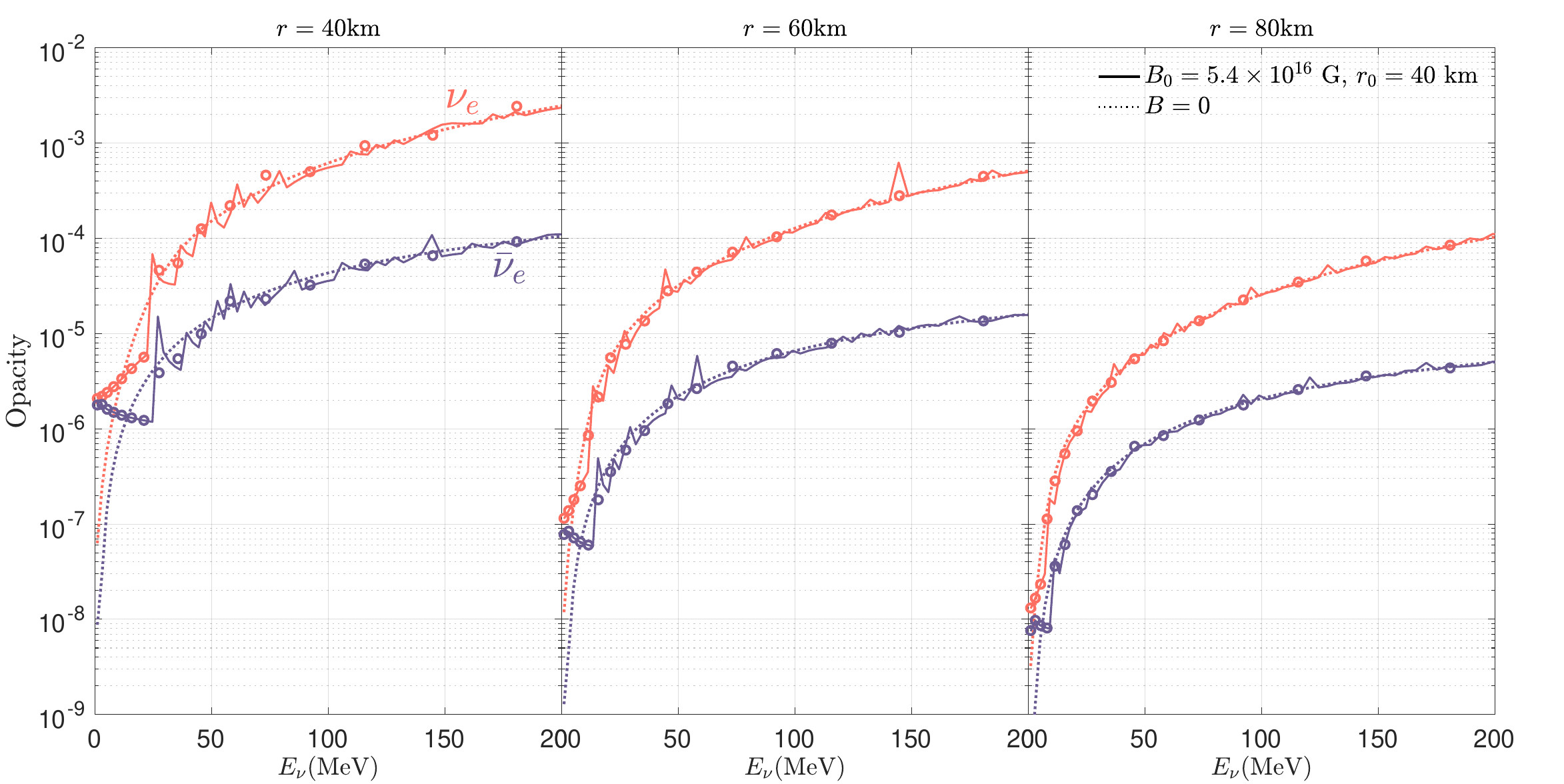}
\caption{\label{opa_comp}Comparison of neutrino opacities between $B_0={5.4}\times 10^{16}$\,G, $r_0=40$\,km model (solid lines) and $B=0$ baseline (dotted-lines). Left, middle and right panel corresponds to $r=40$\,km, 60\,km and 80\,km, respectively. The magnetic field forms a discontinuous opacities compare with the baseline. The simulation uses a 4 point Gauss-Laguerre quadrature weights to calculate opacities to ensure numerical stability, as shown in unfilled circles.}
\end{figure}
{
In Fig. \ref{opa_comp}, we compare the neutrino opacities for 3 typical radii 40\,km, 60\,km, and 80\,km. The values are calculated using the extracted temperature, density and $Y_e$ profiles from the simulation. The dotted and solid lines correspond to the $B=0$ baseline and a dipole magnetic field with $B_0={5.4}\times 10^{16}$\,G, $r_0=40$\,km, respectively. The condition $E_e^2-m_e^2 -2neB>0$ imposes a neutrino energy threshold $E_\nu>20$\,MeV for interactions with electrons occupying $n>1$ Landau levels, generating opacity discontinuities. This quantization mechanism systematically repeats at critical energies corresponding to higher Landau levels. Above $E_\nu>100$\,MeV, opacity profiles converge as many Landau levels become accessible, effectively recovering the $B=0$ limit. These characteristic patterns become unresolvable in the right panel due to the decay of the dipole magnetic field $B\propto 1/r^3$. To avoid discontinuous shapes and ensure numerical stability, the opacity and emissivity calculation utilizes the 4-point Gauss-Laguerre quadrature (indicated by unfilled circles in each panel). This approach effectively smooths discontinuities at high energy, while preserving essential features at $E_\nu<20$\,MeV.
}

Both $\kappa_\nu$ and $\eta_\nu$ directly determine the neutrino emission and absorption source term $S^\alpha_{\rm e/a}$ in Equations (50) and (51) of~\citet{OConnor:2014sgn} and affect the detailed neutrino transport. To verify that our modification is correct, we test a very weak magnetic field (${4.4}\times 10^{12}$\,G) model to ensure that the results are consistent with the $B=0$ case.
We also note that our 1-D simulations do not incorporate magnetic pressure, which would greatly affect the hydrodynamics. However, the magnetic field modifications to neutrino transport in this study are additional effects apart from the hydrodynamics. Conceptually, within the MHD framework, the $B = 0$ case in this study corresponds to a configuration where the magnetic field is present but does not alter neutrino transport.

\section{Results}
\label{results}
\begin{figure}
\centering
\includegraphics[width=1.0\textwidth,clip=true,trim=0cm 0cm 0cm 0cm]{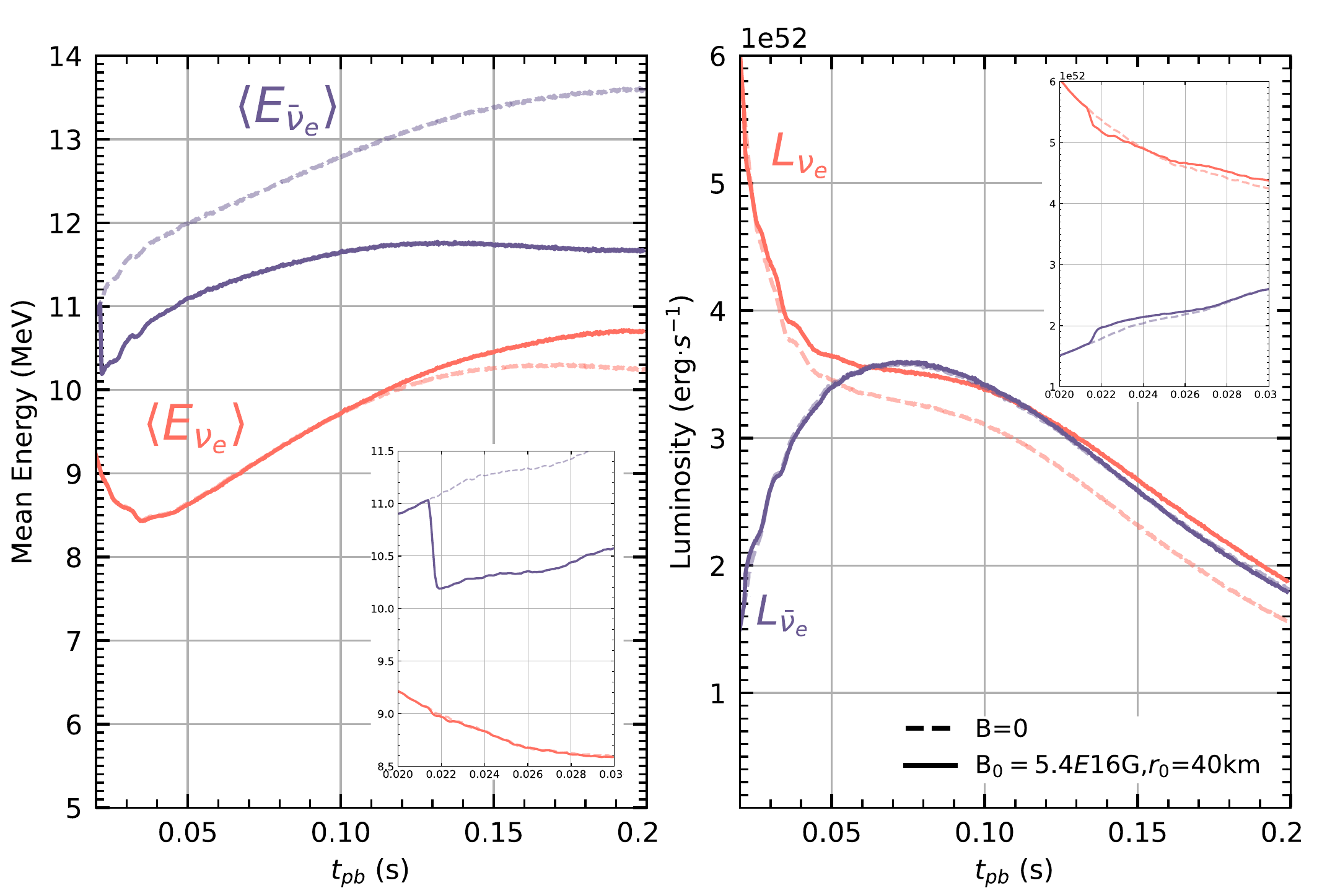}
\caption{\label{r40_simu}Left panel: Comparison of the $\langle E_{{\nu_e}} \rangle$ and $\langle E_{\bar{\nu_e}} \rangle$ evolution between the $B_0={5.4}\times10^{16}$\,G, $r_0=40$\,km model (solid lines) and the $B=0$ baseline (dashed lines). Right panel: Comparison of the $L_{\nu_e}$ and $L_{\bar\nu_e}$ evolution for these two cases. The line-color style follows the left panel. The inset of both panels displays the zoom-in of the initial changes when the magnetic field is just introduced at $t_{pb}\sim 0.02$\,s.}
\end{figure}

To clarify the role of magnetic fields in neutrino transport, we first analyze the model with a dipole magnetic field $B= B_0\cdot r_0^3/r^3$ where $B_0={5.4}\times10^{16}$\,G and $r_0=40$\,km (i.e., magnetic field stays constant for $r<r_0$ and decays proportionally to $1/r^3$ for $r>r_0$) in detail. 
Figure. \ref{r40_simu} shows the mean energy ($\langle E_{\nu_e} \rangle$ and $\langle E_{\bar{\nu_e}} \rangle$) evolution (left panel) and the neutrino luminosity ($L_{\nu_e}$ and $L_{\bar\nu_e}$) evolution (right panel). The solid lines in each panel represent the results for the $B_0={5.4}\times10^{16}$\,G, $r_0=40$\,km model, while the dashed lines represent the $B=0$ non-magnetic baseline.
The magnetic field reduces $\langle E_{\nu_e} \rangle$ and $\langle E_{\bar{\nu_e}} \rangle$ at $t_{pb}\sim 0.02$\,s initially, then at $t_{pb}>0.1$\,s,  $\langle E_{\nu_e} \rangle$ gradually exceeds the value of the $B=0$ scenario. The magnetic field also initially suppresses neutrino luminosity $L_{\nu_e}$, but $L_{\nu_e}$ gradually overtakes the $B=0$ baseline later. Conversely, $L_{\bar\nu_e}$ exhibits a small increase initially and then returns to a level the same as the baseline.

\begin{figure}
\centering
\includegraphics[width=1.0\textwidth,clip=true,trim=0cm 0cm 0cm 0cm]{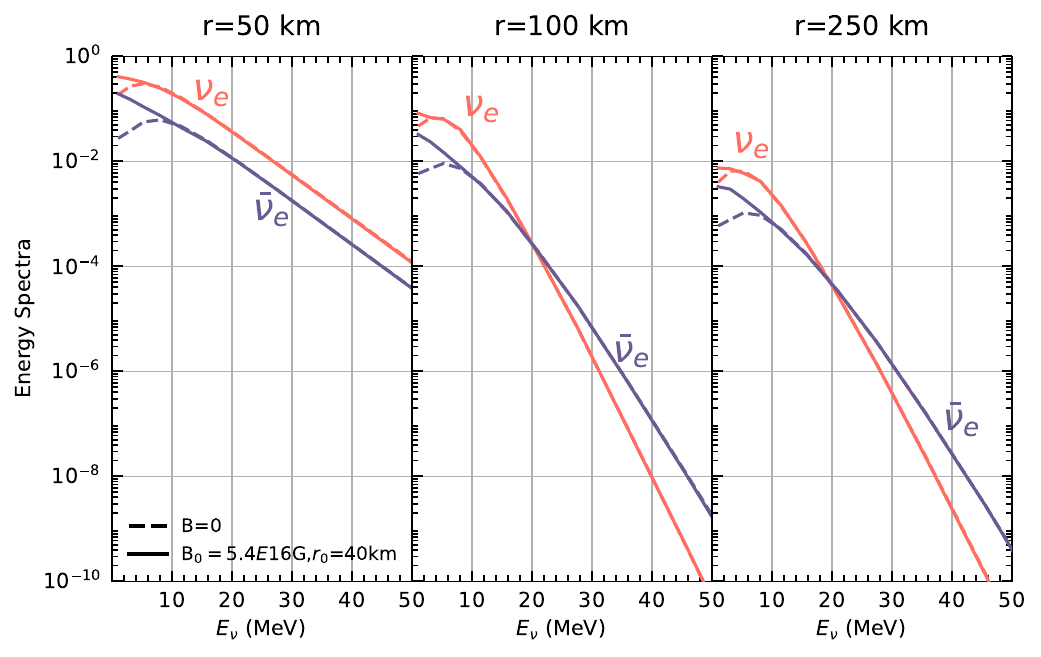}
\caption{\label{ene_spec} 
The neutrino energy spectra at post-bounce time $t_{pb}=0.023$\,s. The spectra are shown for $r = \{50,100,250\}$\,km for left, middle, and right panels, respectively. Magnetic field effects produce low-energy spectral enhancement ($E_\nu<10$\,MeV) via modified cross sections $\sigma_B$ (solid lines). This effect is particularly pronounced for $\bar\nu_e$ due to a lower emitted $e^+$ momentum. A dipole magnetic field with $B_0={5.4}\times10^{16}$\,G and $r_0=40$\,km is applied for the solid curves. Dashed curves show baseline non-magnetic case.}
\end{figure}

For this model, $eB_0=319.87\rm\,MeV^2$, {within $r<40$\,km}, the low-energy neutrinos with $E_{\nu_e}<16.6$\,MeV only interact with $e^-$ that are trapped to the lowest Landau level, i.e., $n=0$ in Eq. \ref{sigma_B}.  Then, the cross section reads
\begin{equation}
    \sigma_B=\frac{G_F^2\cos^2\theta_C(f^2+3g^2)}{\pi}\cdot \frac{eB E_e}{p_e},
\end{equation}
comparing it with the standard expression:
\begin{equation}
    \sigma_0=\frac{G_F^2\cos^2\theta_C(f^2+3g^2)}{\pi} E_e p_e,
\end{equation}
one can find $\sigma_B >\sigma_0$ and the enhance factor is $eB/p_e^2$. This factor is about $20$ for $\nu_e$ with $E_{\nu}<10$\,MeV. However, for $\bar\nu_e$, $eB/p_e^2>100$ since for the same incident neutrino energy, the emitted $e^+$ has a smaller $p_e$ value due to the mass difference between proton and neutron. This enlarged cross section directly affects the energy spectra for $\nu_e$ and $\bar\nu_e$. 
Figure. \ref{ene_spec} presents the $\nu_e$ and $\bar\nu_e$ energy spectra at $t_{pb}=0.023$\,s for the region $r=50$\,km (left panel), $r=100$\,km (middle panel) and $r=250$\,km (right panel), respectively. Solid lines are the energy spectra taken from the $B_0={5.4}\times10^{16}$\,G, $r_0=40$\,km model, which exhibits an enhanced neutrino spectrum at $E_\nu<10$\,MeV. A more pronounced impact on $\bar\nu_e$ (blue lines) compared to $\nu_e$ (red lines) is observed across all panels, attributable to the lower momentum of emitted $e^+$. The energetic neutrinos with $E_\nu>20$\,MeV allow more electrons Landau levels to pop-up, so $\sigma_B$ becomes identical to $\sigma_0$, resulting in the indistinguishable high-energy spectra. Such lower energy weighted spectra from the $\sigma_B$ modification further leads to the initial mean energy shift in the left panel of Fig. \ref{r40_simu}, i.e., $\Delta\langle E_{\bar\nu_e} \rangle \approx -1$\,MeV and $\Delta\langle E_{\nu_e} \rangle \approx -0.1$\,MeV, compared with the $B=0$ baseline.

\begin{figure}
\centering
\includegraphics[width=1.0\textwidth,clip=true,trim=0cm 0cm 0cm 0cm]{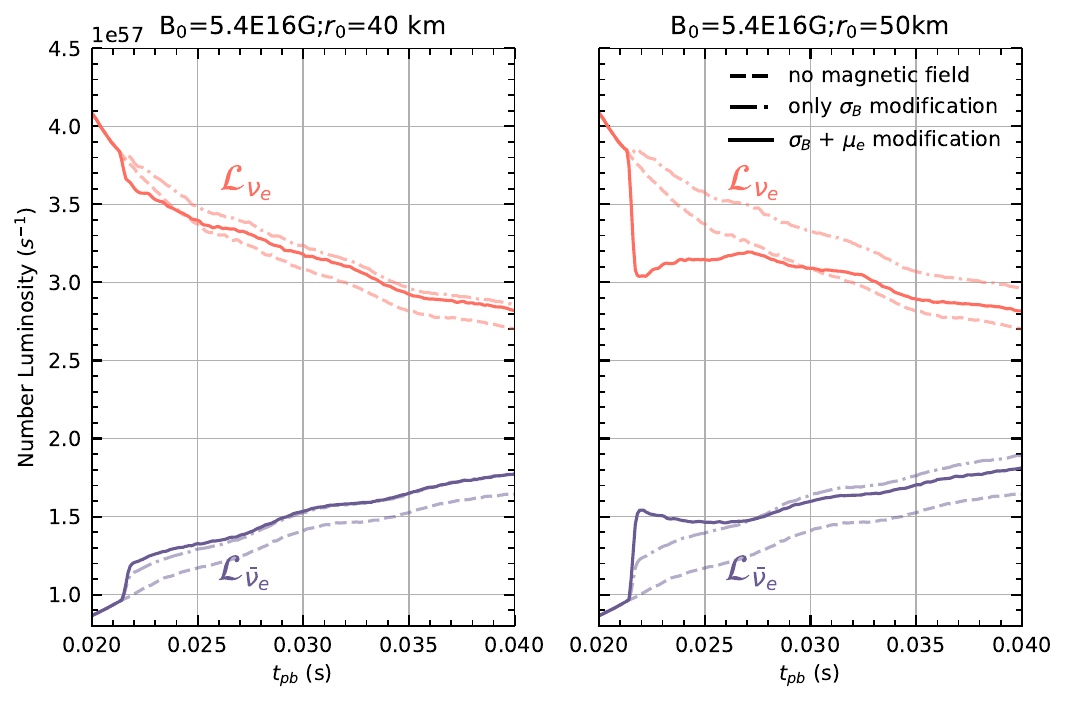}
\caption{\label{numlum_evol}Left panel: The neutrino number luminosity $\mathcal{L}_{\nu_e}$ and $\mathcal{L}_{\bar{\nu}_e}$ evolution for the model $B_0={5.4}\times10^{16}\,{\rm G} ,r_0=40$\,km. Right panel: $\mathcal{L}_{\nu_e}$ and $\mathcal{L}_{\bar{\nu}_e}$ evolution for the model $B_0={5.4}\times10^{16}\,{\rm G} ,r_0=50$\,km. With modified cross sections $\sigma_B$ (dash-dotted lines), both $\mathcal{L}_{\nu_e}$ and $\mathcal{L}_{\bar{\nu}_e}$ show slightly enhancement compare with the $B=0$ baseline (dashed lines). For the full magnetic field impact, i.e., $\sigma_B$ and $\mu_e$ modification (solid lines),  $\mathcal{L}_{\nu_e}$ demonstrates a suppression while $\mathcal{L}_{\bar\nu_e}$ amplifies.}
\end{figure}

Figure. \ref{numlum_evol} shows the evolution of number luminosity ($\mathcal{L}_{\nu_e}$ and $\mathcal{L}_{\bar\nu_e}$) during $t_{pb} = 0.02-0.04\,\mathrm{s}$ for models with $B_0={5.4}\times10^{16}\,{\rm G} ,r_0=40$\,km (left panel) and $B_0={5.4}\times10^{16}\,{\rm G} ,r_0=50$\,km (right panel), separately.
Their behavior stems from two factors: one is the lower energy weighted spectrum caused by $\sigma_B$, {this results in enhancements of  $\mathcal{L_\nu}$, as shown in dot-dashed lines of each panel.} The other factor is the suppression of $\mu_e$ due to the strong magnetic field. If $eB>\pi^{3/2}[2(\rho Y_e)^2]^{1/3}$, electrons only occupy the lowest Landau level and have a smaller {value of} $\mu_e$ \cite{Luo:2024qmq}. 
This suppression propagates to neutrino chemical potentials via $\mu_{\nu_e} = \mu_e - (\mu_n - \mu_p) = -\mu_{\bar{\nu}_e}$ in the equilibrium state, thus diminishing the distribution function term in black body spectrum $B_{\nu_e}$ while amplifying that of $B_{\bar{\nu}_e}$, {resulting in a decreased $\mathcal{L}_{\nu_e}$ and an increased $\mathcal{L}_{\bar\nu_e}$, as shown in solid lines.}

{Moreover}, the extra $\mu_e$  modification attributes to a `perturbation' of the number luminosity (solid lines): at the moment that magnetic field is introduced, $\mathcal{L}_{\bar{\nu}_e}$ rises by an additional ${4}\%$, whereas $\mathcal{L}_{\nu_e}$ declines by {the same level} on the left panel. These impacts are more significant in the right panel, where $\mathcal{L}_{\bar{\nu}_e}$ rises by ${30}\%$, while $\mathcal{L}_{\nu_e}$ is hampered by ${20}\%$. 
Following this abrupt change, $\mathcal{L}_{\nu_e}$ gradually increases and $\mathcal{L}_{\bar\nu_e}$ decreases. By $t_{pb}\sim0.04$\,s, $\mathcal{L}_{\nu_e}$ and $\mathcal{L}_{\bar\nu_e}$ recover to the level of the `only $\sigma_B$ modification' scenario. This `perturbation' behavior is explained by the evolution of $Y_e$, as discussed below.

\begin{figure}
\centering
\includegraphics[width=1.0\textwidth,clip=true,trim=0cm 0cm 0cm 0cm]{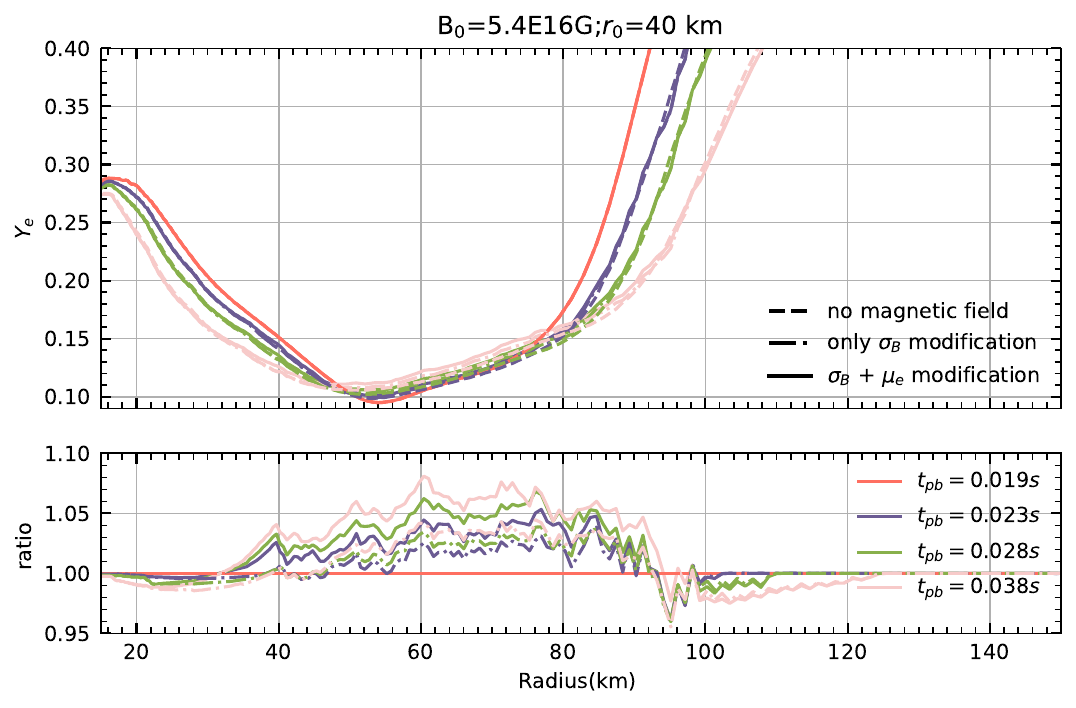}
\caption{\label{Ye_evol}$Y_e$ of the stellar matter when $0.02<t_{pb}<0.04$\,s. The ratio compared to the $B=0$ baseline is shown in the lower panel for both $\sigma_B$ modification model (dash-dotted lines) and full magnetic field modification model (solid lines).}
\end{figure}

The unbalanced magnetic field impact on $\nu_e$ and $\bar\nu_e$ drives the proton enrichment, as shown in Fig. \ref{Ye_evol}. This figure compares the value of $Y_e$ at $0.02<t_{pb}<0.04$\,s for the $B=0$ baseline (dashed lines), the case `only $\sigma_B$ modification' (dashed-dotted lines), and the full magnetic impact (solid lines). The ratio to the baseline is shown in the lower panel. A $3\%$ enhancement $Y_e$ in the region $30-100$\,km is due to the $\sigma_B$ modification, the additional $\mu_e$ modification further contributes to an extra $5\%$ in the same region. The rise of $Y_e$, equivalent to an enlarged electron number density, gradually reinforces the $\nu_e$ emission due to an increased $e^-+p\to n+ \nu_e$ rate, but hampers the $\bar\nu_e$ emission. These combined effects directly generate the $\mathcal{L}_{\nu}$ `perturbation' as shown in Fig. \ref{numlum_evol}.

\begin{figure}
\centering
\includegraphics[width=1.0\textwidth,clip=true,trim=0cm 0cm 0cm 0cm]{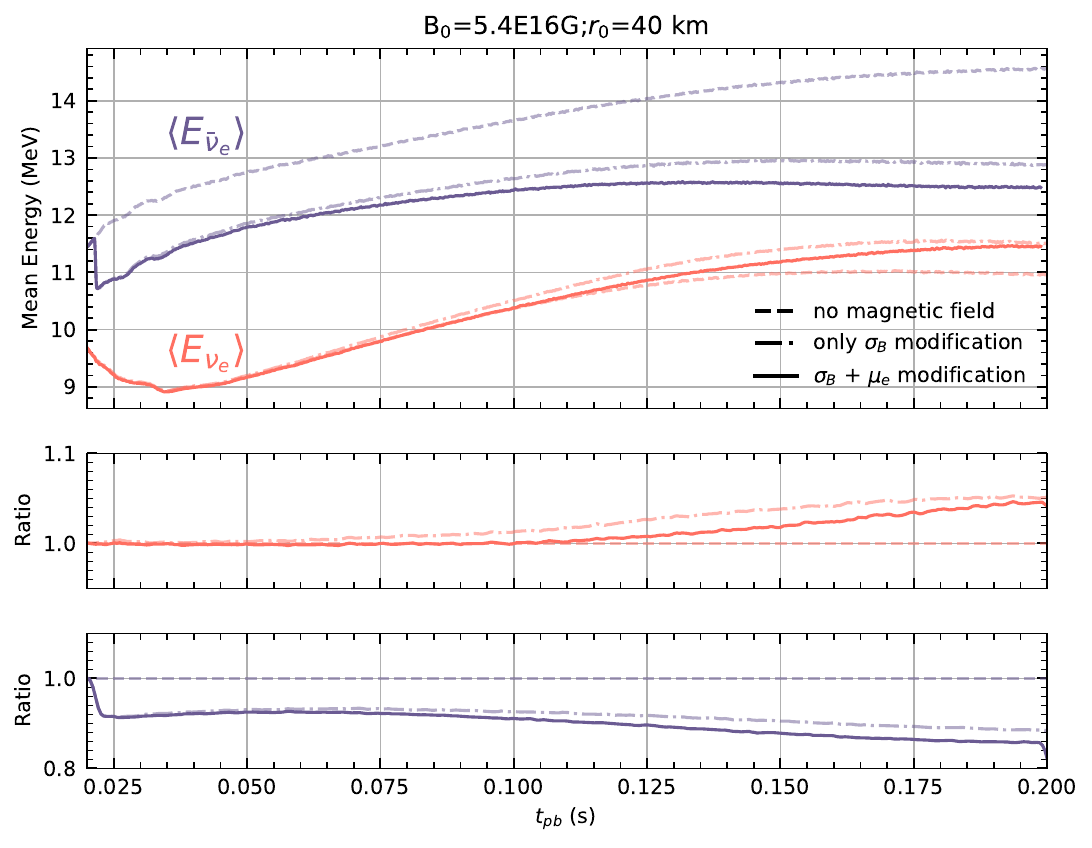}
\caption{\label{energy_comp}Upper panel: $\langle E_{{\nu_e}} \rangle$ and $\langle E_{\bar{\nu_e}} \rangle$ evolution for $t_{pb}=0.02$--0.2\,s. 
With modified cross sections $\sigma_B$, $\langle E_{\bar{\nu_e}} \rangle$ decreases due to the lower energy weighted spectrum (dash-dotted blue line). The further enhanced $Y_e$ at $30-100$\,km gradually increases $\langle E_{{\nu_e}} \rangle$ (dash-dotted red line). The additional $\mu_e$ modification further suppresses $\langle E_{{\nu_e}} \rangle$ and $\langle E_{\bar{\nu_e}} \rangle$ (solid lines), yielding the net shifts of $\Delta \langle E_{\nu_e} \rangle=0.5$\,MeV) while $\Delta \langle E_{\bar{\nu_e}} \rangle = -2$\,MeV at $t_{pb}=0.2$\,s, compared with $B=0$ baseline (dashed lines). Middle panel: $\langle E_{\bar{\nu_e}} \rangle$ ratio for $\sigma_B$ modification case and full modification case relative to the $B=0$ baseline. Lower panel: $\langle E_{{\nu_e}} \rangle$ ratio for $\sigma_B$ modification case and full modification case relative to the $B=0$ baseline. }
\end{figure}

The enlarged $Y_e$ further affects the neutrino mean energy and we compare its evolution for $t_{pb}=0.02$--0.2\,s in Fig. \ref{energy_comp}. The $Y_e$ at 30--100\,km leads to an enhanced $\langle E_{\nu_e} \rangle$ (dashed-dotted blue line) and reduced $\langle E_{\bar{\nu_e}} \rangle$ (dashed-dotted red line), since a higher electron density results in a greater value of $\mu_e$. After adding the $\mu_e$ correction stemming from the magnetic field, the $Y_e$ is further enlarged. Although this effect would further enhance $\langle E_{{\nu_e}} \rangle$ and reduce $\langle E_{\bar{\nu_e}} \rangle$, the direct suppression of $\mu_e$ value due to the magnetic field partially offsets these trends. As a result, values of $\langle E_{{\nu_e}} \rangle$ and $\langle E_{{\bar\nu_e}} \rangle$ (solid lines) are slightly suppressed compared to that of `only $\sigma_B$ modification' case (dashed-dotted lines), yielding the net shifts of $\Delta \langle E_{\nu_e} \rangle=0.5$\,MeV and $\Delta \langle E_{\bar{\nu_e}} \rangle = -2$\,MeV at $t_{pb}=0.2$\,s.

In summary, in the early phase, the enlarged cross section of the low-energy neutrinos leads to a lower energy weighted spectrum, thereby reducing the mean neutrino energy and enlarging their number luminosity. The $e^+$ has a smaller momentum due to the mass difference between proton and neutron, which amplifies these impacts for $\bar\nu_e$ compared with $\nu_e$. Furthermore, the suppressed $\mu_e$ under magnetic fields modifies the black body spectrum, decreasing the distribution function term of $B_{\nu_e}$ while increasing that of $B_{\bar\nu_e}$. 
Although it acts like a `perturbation' to the number luminosity, the resultant unbalanced neutrino emission elevated $Y_e$ value at $r=30$--100\,km for about $8\%$, equivalent to an enlarged electron number density. Therefore, $\langle E_{\bar{\nu_e}} \rangle$ decreases and $\langle E_{{\nu_e}} \rangle$ increases. On the other hand, $\mu_e$ suppression due to the magnetic field partially counteracts these trends, resulting in $\Delta\langle E_{\nu_e} \rangle=0.5$\,MeV while $\Delta\langle E_{\bar{\nu_e}} \rangle=-2$\,MeV at $t_{pb}=0.2$\,s. Finally, the enlarged $\mathcal{L}_{{\nu}_e}$and $\langle E_{{\nu_e}} \rangle $ enhances $L_{\nu_e}$, while the suppressed $\langle E_{\bar{\nu_e}} \rangle $ counterbalances the elevated $\mathcal{L}_{\bar{\nu}_e}$, hence $L_{\bar\nu_e}$ slightly decreases, as seen in Fig. \ref{r40_simu}. 
{These effects complement the previous work: \citet{Duan:2004nc, Duan:2005fc} analyzed heating and cooling rates under magnetic fields using an analytical approach without real simulations, while \citet{2021ApJ...906..128K} includes the modified neutrino-nucleus cross section in a 3D MHD framework and investigates its impact on the asymmetric explosion patterns. However, neither study accounted for electron chemical potential corrections. Our results incorporate both cross section modifications and chemical potential corrections, showing the influence of magnetic fields on neutrino luminosity, neutrino mean energy, and electron fraction. These quantities are critical for CCSN nucleosynthesis which indicate possible imprints of the magnetic field in the abundance pattern.} 

\begin{figure}
\centering
\includegraphics[width=1.0\textwidth,clip=true,trim=0cm 0cm 0cm 0cm]{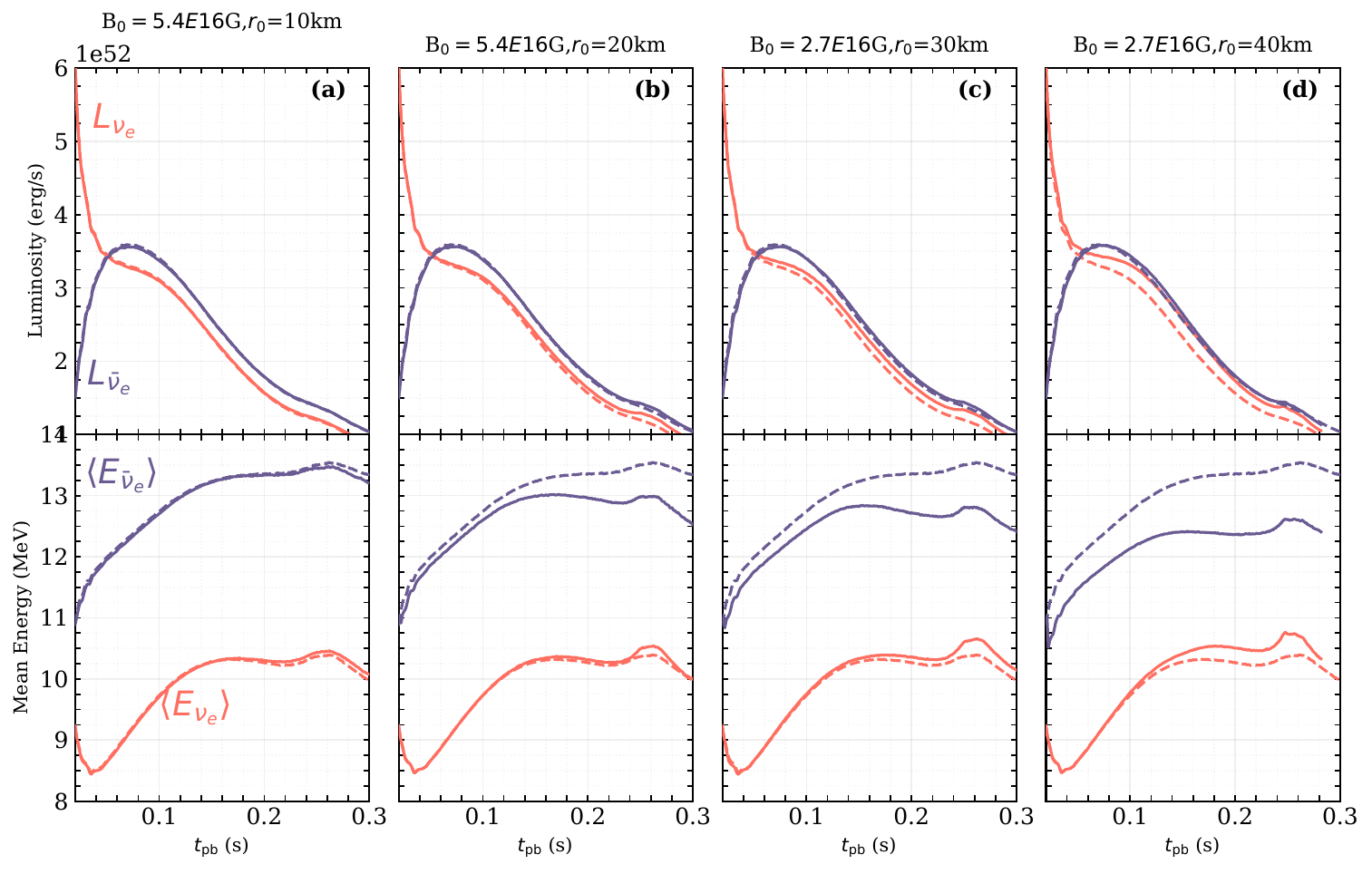}
\includegraphics[width=1.0\textwidth,clip=true,trim=0cm 0cm 0cm 0cm]{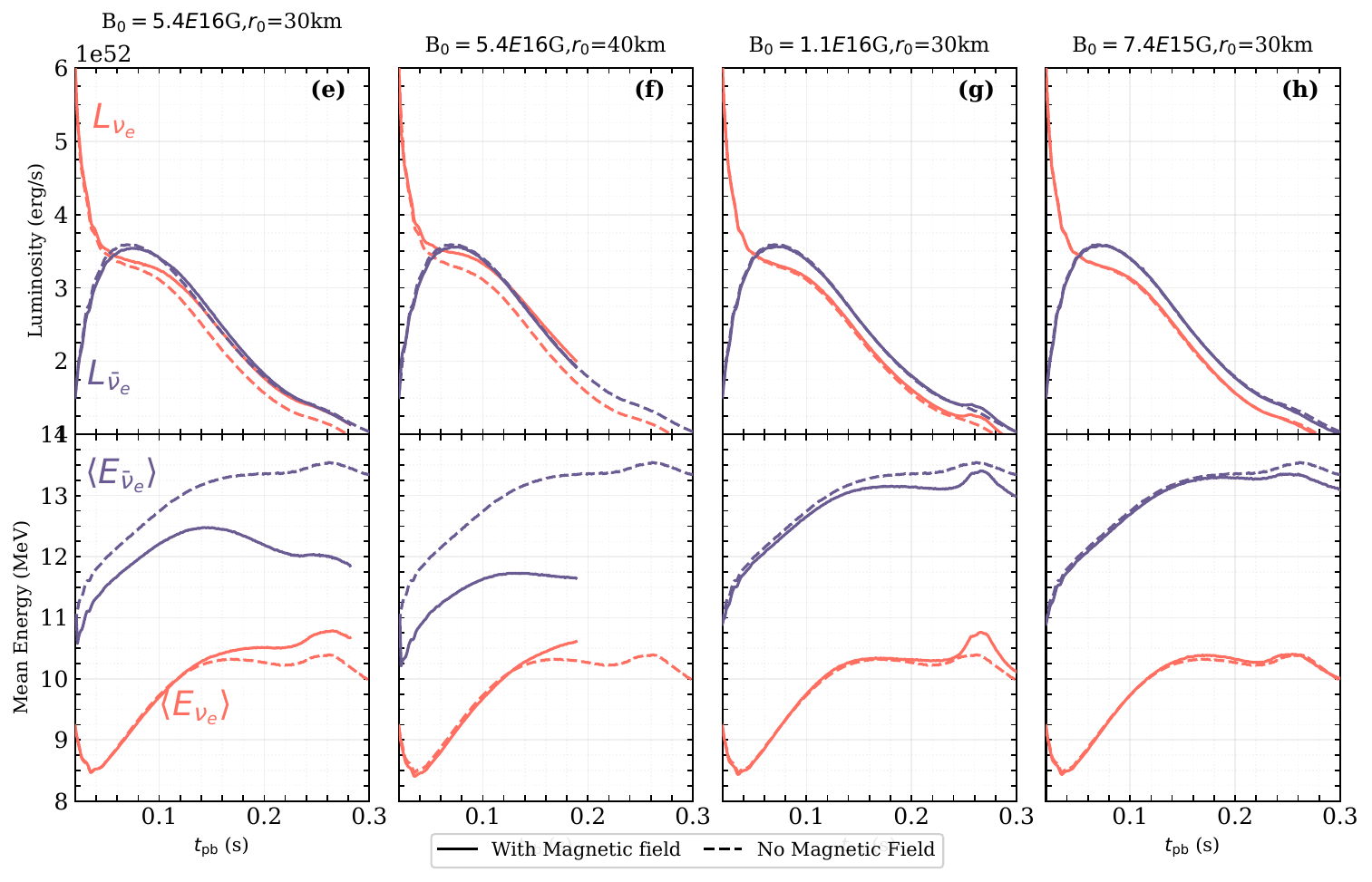}
\caption{\label{simu_results} $\langle E_{\bar{\nu}} \rangle$ and $L_\nu$ evolution from simulations with different ($B_0$, $r_0$) sets. The line-color style follows Fig. \ref{r40_simu}. Notice that in panel (f), the simulation stops at $t_{pb}\sim$0.2\,s because the strong magnetic field elevates $Y_e$ beyond the maximum value supported by the EoS table.}
\end{figure}

The similar analysis can be applied beyond this specific model. We carry on a systematic study by running the simulation for various $(B_0,r_0)$ sets. The corresponding simulated neutrino luminosity $L_\nu$ and neutrino mean energy $\langle E_\nu \rangle$ are shown in Fig. \ref{simu_results}.

Results with and without magnetic fields are plotted in solid and dashed lines in each panel, respectively. For $r_0<20$\,km, even with a strong magnetic field of $B_0= {5.4}\times 10^{16}$\,G, there is no significant deviation from the $B=0$ baseline. Only $\langle E_{\bar{\nu}_e} \rangle$ exhibited a small reduction ($\sim$0.75\,MeV) at $t_{\rm pb} > 0.2$ s, as shown in panels (a) and (b). 
The neutrino mean energy $\langle E_{\bar{\nu}_e} \rangle$ demonstrates a strong dependence on $r_0$. When $r_0$ increases to 30--40\,km, $\Delta\langle E_{\bar{\nu}_e} \rangle$ reaches 0.75--1\,MeV for $B_0 = {2.7} \times 10^{16}$\,G (panel (c) and (d)) and exceeds $1.25$\,MeV for $B_0= {5.4}\times 10^{16}$\,G (panel (e) and (f)). $L_{\nu_e}$ is enhanced by 8--10\% in the $B_0= {5.4}\times 10^{16}$\,G, $r_0=30$--40\,km models while $L_{\bar{\nu}_e}$ shows a negligible magnetic field dependence. This is because the increased $\mathcal{L}_{\bar{\nu}_e}$ is counterbalanced by the suppressed mean energy $\langle E_{\bar{\nu}_e} \rangle$. $\langle E_{\bar{\nu}} \rangle$ and $L_\nu$ are crucial for the later nucleosynthesis such as $\nu$p-process and $\nu$-process. Their strong dependence on the magnetic field indicates possible imprints of the magnetic field in the abundance pattern of CCSN nucleosynthesis. The threshold below which the above impacts become negligible is found to be $B_0\sim{7.4}\times10^{15}$\,G--${1.1}\times10^{16}$\,G when $r_0=$30--40\,km, as both panels (g) and (h) show only minimal deviations in the mean neutrino energy.

\section{Conclusion}\label{con_dis}
In this study, we investigated the impact of strong magnetic fields on the neutrino transport inside core-collapse supernovae. The magnetic field enlarges the cross sections of the neutrino absorption rate for low-energy $\nu_e$ and $\bar\nu_e$. It suppresses the chemical potential of $e^\pm$ since they only occupy the lowest Landau level under a strong magnetic field. 
We included these two major effects in the M1 scheme for neutrino transport and employed the \texttt{GR1D} code to perform 1-D CCSN simulations of a 9.6 M$_{\odot}$ zero-metallicity progenitor. 
The enlarged low-energy cross section reduces the $\bar\nu_e$ mean energy $\langle E_{\bar{\nu}_e} \rangle$ and increases the number luminosities for both $\nu_e$ and $\bar\nu_e$. At the same time, an additional $\mu_e$ modification under strong magnetic fields enhances $\bar\nu_e$ emission but hampers that of $\nu_e$. These combined effects increase $Y_e$ at $30-100$\,km, further enlarging $\nu_e$ mean energy due to a larger electron number density.
As for the neutrino luminosity, an enlarged number luminosity and mean energy amplify $L_{\nu_e}$ compared with the non-magnetic case. On the other hand, the elevated $\mathcal{L}_{\bar{\nu}_e}$ counterbalances the suppressed $\langle E_{\bar\nu_e}\rangle$, resulting in a minimal modification to $L_{\bar\nu_e}$.

We conducted a systematic study by simulating various dipole magnetic field ($B= B_0\cdot r_0^3/r^3$) models. For $r_0<20$\,km, fields up to $B_0= {5.4}\times 10^{16}$\,G cause negligible deviations from the non-magnetic case, except for minor $\langle E_{\bar{\nu}_e} \rangle$ reductions. If $r_0>30$\,km,  $\langle E_{\bar{\nu}_e} \rangle$ becomes significantly suppressed and $L_{{\nu}_e}$  shows substantial enhancement
for $B_0 \geq {2.7}  \times 10^{16}$\,G. These effects are negligible if $B_0<{7.4}\times10^{15}$\,G, otherwise the magnetic-field-dependent behaviors of $\langle E_{\bar{\nu}_e} \rangle$ and $L_\nu$ may leave imprints in the CCSN nucleosynthetic abundance patterns.

\begin{acknowledgments}
We thank Motohiko Kusakabe, Ko Nakamura, Nobuya Nishimura, and Kanji Mori for their valuable discussions. Y. Luo thanks the Institute of Modern Physics, Tsung-Dao Lee Institute at Shanghai Jiao Tong University, and Nanjing University during his visits. This work is supported by the National Natural Science Foundation of China (No. 12335009 \& 12435010) and National Key R\&D Program of China (2022YFA1602401). Y. L. is supported by the China Postdoctoral Science Foundation No. 2025T180924 and the Boya Fellowship of Peking University. S. Zha is supported by the National Natural Science Foundation of China (NSFC, Nos. 12288102, 12393811, 12473031), the International Centre of Supernovae, Yunnan Key Laboratory (No. 202505AV340004), the Yunnan Fundamental Research Project (Nos. 202401BC070007, 202501AS070078) and the Yunnan Revitalization Talent Support Program--Young Talent project. T. Kajino is supported in part by Grants-in-Aid for Scientific Research of Japan Society for the Promotion of Science (20K03958).
\end{acknowledgments}

\appendix

\section{Weak Interactions in Strong Magnetic Field}\label{B_impact}
\subsection{Interaction cross section}\label{Weak_B}
This work only considers magnetic field strength smaller than $10^{18}$\,G, which cannot affect weak interaction Lagrangian. Since the nuclear magneton is three orders of magnitude smaller than the Bohr magneton due to heavy nuclear mass, the magnetic effects on nucleons are ignored. Three active neutrinos are also free from the magnetic field effects. The cross section is given by the integration of $d\sigma$, whose expression is~\cite{Peskin:1995ev}:
\begin{eqnarray}
&&d\sigma = \sum_{s_p}\sum_{s_e}\dfrac{Ldk_{\rm ez}}{ (2\pi)}\dfrac{1}{ 2E_e 2E_p}\dfrac{L^3 d^3k_p}{(2\pi)^3}\dfrac{|iM|^2}{ 2E_{\nu_e}2E_n|v_{\nu_e}-v_n|}\times\nonumber\\
&&  (2\pi)^2\delta(E_e+E_p-E_\nu-E_n) \delta(p_e+p_p-p_\nu-p_n),
\end{eqnarray}
where $M$ is the interaction amplitude. We applied the results from~\citet{Duan:2004nc,Duan:2005fc} with the $0$-th order of approximation:
\begin{eqnarray}
\label{sigma_B}
    \sigma_{\rm \nu N}(E_n,B) &=& \sigma_B^1\Big[1+2\chi \dfrac{(f\pm g)g }{ f^2 +3g^2} \cos \Theta_\nu\Big]+\nonumber  \\
    & \sigma_B^2&\Big[\dfrac{f^2 -g^2 }{ f^2 +3g^2}\cos\Theta_\nu + 2\chi \dfrac{(f\mp g) g }{ f^2 + 3g^2} \Big],
\end{eqnarray}
where we have
\begin{eqnarray}
    &&\sigma_B^1 = \dfrac{G_F^2 \cos^2\theta_C }{ 2\pi}(f^2 + 3g^2)\times \nonumber\\
    &&eB\sum^{n_{\rm max}}_{n=0}\sum^{s=1}_{s=-1}\dfrac{g_nE_n }{ \sqrt{E_n^2 - m_e^2 -(2n+s-1)eB}},
\end{eqnarray}
and
\begin{equation}
\sigma_B^2 = \dfrac{G_F^2 \cos^2\theta_C }{ 2\pi}(f^2 + 3g^2)eB\dfrac{E_n}{ \sqrt{E_n^2 - m_e^2}},  
\end{equation}
where $E_n$ is given by
\begin{equation}
    E_n^2 = p_z^2+m_e^2+ (2n+s-1)eB,
\end{equation}
$n$ is the Landau quantum number~\cite{Landau:1991wop} and $s=\pm1$ is the spin of electrons and positrons. In Eq.~\ref{sigma_B}, $\Theta_\nu $ is the angle between the neutrino momentum and the magnetic field, the upper sign is for $\nu_e +n$ reaction, and the lower sign is for $\bar{\nu}_e + p$. In~\citet{Duan:2005fc}, higher-order terms are considered to correct the nucleon mass effect during interaction. However, we ignore this correction since it does not affect the cross section significantly (see Fig. 3 of~\citet{Duan:2005fc}). Several previous studies~\cite{Maruyama:2021ghf,Famiano:2022lmz,Famiano:2020fbq,Luo:2020slj} have also adopted the same approximation. 

\subsection{Chemical potential}\label{effect_chemi}
The modification of weak interaction rates inside a magnetic field has been discussed in several papers~\cite{Duan:2004nc,Lai:1998sz,Arras:1998mv}. However, inside the magnetic field, $e^\pm$ obey the Fermi-Dirac distribution but with the Landau quantization of $E_n$ as we mentioned above, so that the chemical potential $\mu_e$ also needs a correction. $\mu_e$ is evaluated from the charge neutrality of plasma, i.e., the net number density of both electrons and positrons should be balanced with the ion. Under a magnetic field, the net electron number density is given by
\begin{eqnarray}
\label{chemi_po}
    \rho N_A Y_e = &n_e&= \frac{m_e\omega_c}{(2\pi)^2}\sum_0^{n_{\rm max}} \int d\bfp \Big[f_{\rm FD}(E_n, \mu_e(B); T)_{s=-1}\nonumber\\
    &+&f_{\rm FD}(E_n, \mu_e(B); T)_{s=1}
        - f_{\rm FD}(E_n, -\mu_e(B); T)_{s=-1}\nonumber\\
        &-&f_{\rm FD}(E_n, -\mu_e(B); T)_{s=1}\Big].
\end{eqnarray}
Here, $N_A$ is the Avogadro number, $\omega_c = eB/m_e$ is the cyclotron frequency, and $s=\pm1$ represents the spin state of electrons and positrons. Notice that inside the magnetic field, the spin is only parallel or anti-parallel with the magnetic field. For the plasma with known electron number density $n_e$, temperature $T$, and field strength $B$, this relation should provide the value of chemical potential. In the limit of $T\to0 $, all electrons are degenerate (there are no positrons since their chemical potential is negative), and the number density is given by:
\begin{equation}
\label{chemi_EF}
    n_e=\frac{2m\omega_c}{(2\pi)^2}\Big[(E_F^2-m_e^2)^{1/2}+2\sum_{n=1}^{n_{\rm max}}\sqrt{E_F^2-m_e^2-2neB}\Big],
\end{equation}
which is consistent with~\citet{DELSANTE1980135}.

\bibliography{B_field_SNe}

@article{Luo:2024qmq,
    author = "Luo, Yudong and Zha, Shuai and Kajino, Toshitaka",
    title = "{Strong magnetic field inside degenerate relativistic plasma and the impacts on the neutrino transport in Core-Collapse Supernovae}",
    journal = "",
    eprint = "2405.11555",
    archivePrefix = "arXiv",
    primaryClass = "astro-ph.HE",
    month = "5",
    year = "2024"
}

@article{Vogel:1983hi,
    author = "Vogel, P.",
    title = "{Analysis of the Anti-neutrino Capture on Protons}",
    reportNumber = "CERN-TH-3727",
    doi = "10.1103/PhysRevD.29.1918",
    journal = "Phys. Rev. D",
    volume = "29",
    pages = "1918",
    year = "1984"
}

@ARTICLE{2002RvMP...74.1015W,
       author = {{Woosley}, S.~E. and {Heger}, A. and {Weaver}, T.~A.},
        title = "{The evolution and explosion of massive stars}",
      journal = {Rev. Mod. Phys.},
         year = 2002,
        month = nov,
       volume = {74},
       number = {4},
        pages = {1015-1071},
          doi = {10.1103/RevModPhys.74.1015},
       adsurl = {https://ui.adsabs.harvard.edu/abs/2002RvMP...74.1015W}
}

@article{takiwaki09,
    title = {Special Relativistic Simulations of Magnetically-dominated Jets in Collapsing Massive Stars},
    volume = {691},
    doi = {10.1088/0004-637X/691/2/1360},
    number = {2},
    journal = {Astrophys. J.},
    author = {Takiwaki, Tomoya and Kotake, Kei and Sato, Katsuhiko},
    year = {2009},
    eprint = "0712.1949",
    archivePrefix = "arXiv",
    pages = {1360--1379},
}

@article{Famiano:2022lmz,
    author = "Famiano, Michael A. and Mathews, Grant and Balantekin, A. Baha and Kajino, Toshitaka and Kusakabe, Motohiko and Mori, Kanji",
    title = "{Evolution of Urca Pairs in the Crusts of Highly Magnetized Neutron Stars}",
    eprint = "2208.09573",
    archivePrefix = "arXiv",
    primaryClass = "astro-ph.HE",
    doi = "10.3847/1538-4357/ac9bf3",
    journal = "Astrophys. J.",
    volume = "940",
    number = "2",
    pages = "108",
    year = "2022"
}

@article{Luo:2020slj,
    author = "Luo, Yudong and Famiano, Michael A. and Kajino, Toshitaka and Kusakabe, Motohiko and Balantekin, A. Baha",
    title = "{Screening corrections to Electron Capture Rates and resulting constraints on Primordial Magnetic Fields}",
    eprint = "2002.08636",
    archivePrefix = "arXiv",
    primaryClass = "astro-ph.CO",
    doi = "10.1103/PhysRevD.101.083010",
    journal = "Phys. Rev. D",
    volume = "101",
    number = "8",
    pages = "083010",
    year = "2020"
}

@article{kiuchi14,
  title = {High resolution numerical relativity simulations for the merger of binary magnetized neutron stars},
  author = {Kiuchi, Kenta and Kyutoku, Koutarou and Sekiguchi, Yuichiro and Shibata, Masaru and Wada, Tomohide},
  journal = {Phys. Rev. D},
  volume = {90},
  issue = {4},
  pages = {041502},
  numpages = {5},
  year = {2014},
  month = {Aug},
  publisher = {American Physical Society},
  doi = {10.1103/PhysRevD.90.041502},
  url = {https://link.aps.org/doi/10.1103/PhysRevD.90.041502}
}

@article{kiuchi15,
  title = {Efficient magnetic-field amplification due to the Kelvin-Helmholtz instability in binary neutron star mergers},
  author = {Kiuchi, Kenta and Cerd\'a-Dur\'an, Pablo and Kyutoku, Koutarou and Sekiguchi, Yuichiro and Shibata, Masaru},
  journal = {Phys. Rev. D},
  volume = {92},
  issue = {12},
  pages = {124034},
  numpages = {11},
  year = {2015},
  month = {Dec},
  publisher = {American Physical Society},
  doi = {10.1103/PhysRevD.92.124034},
  url = {https://link.aps.org/doi/10.1103/PhysRevD.92.124034}
}

@article{nakamura15,
    title = {r-process nucleosynthesis in the {MHD}+neutrino-heated collapsar jet},
    volume = {582},
    doi = {10.1051/0004-6361/201526110},
    journal = {Astron. Astrophys.},
    author = {Nakamura, K. and Kajino, T. and Mathews, G. J. and Sato, S. and Harikae, S.},
    year = {2015},
    pages = {A34}
}

@ARTICLE{ruiz,
       author = {{Ruiz}, Milton and {Tsokaros}, Antonios and {Shapiro}, Stuart L.},
        title = "{Magnetohydrodynamic simulations of binary neutron star mergers in general relativity: Effects of magnetic field orientation on jet launching}",
      journal = {Phys. Rev. D},
         year = 2020,
       volume = {101},
       number = {6},
          eid = {064042},
        pages = {064042},
          doi = {10.1103/PhysRevD.101.064042},
archivePrefix = {arXiv},
       eprint = {2001.09153},
 primaryClass = {astro-ph.HE},
       adsurl = {https://ui.adsabs.harvard.edu/abs/2020PhRvD.101f4042R}
}

@article{price,
    author = "Price, Daniel and Rosswog, Stephan",
    title = "{Producing ultra-strong magnetic fields in neutron star mergers}",
    eprint = "astro-ph/0603845",
    archivePrefix = "arXiv",
    doi = "10.1126/science.1125201",
    journal = "Science",
    volume = "312",
    pages = "719",
    year = "2006"
}

@article{DELSANTE1980135,
title = {Dielectric response of a relativistic degenerate electron plasma in a strong magnetic field},
journal = {Annals of Physics},
volume = {125},
number = {1},
pages = {135-175},
year = {1980},
doi = {https://doi.org/10.1016/0003-4916(80)90122-0},
author = {A.E. Delsante and N.E. Frankel}
}

@article{Arras:1998mv,
    author = "Arras, Phil and Lai, Dong",
    title = "{Neutrino - nucleon interactions in magnetized neutron star matter: The Effects of parity violation}",
    eprint = "astro-ph/9811371",
    archivePrefix = "arXiv",
    doi = "10.1103/PhysRevD.60.043001",
    journal = "Phys. Rev. D",
    volume = "60",
    pages = "043001",
    year = "1999"
}

@article{Lai:1998sz,
    author = "Lai, Dong and Qian, Yong-Zhong",
    title = "{Neutrino transport in strongly magnetized proto neutron stars and the origin of pulsar kicks. 2. The Effect of asymmetric magnetic field topology}",
    eprint = "astro-ph/9802345",
    archivePrefix = "arXiv",
    doi = "10.1086/306203",
    journal = "Astrophys. J.",
    volume = "505",
    pages = "844",
    year = "1998"
}

@book{Landau:1991wop,
    author = "Landau, Lev Davidovich and Lifshits, E. M.",
    title = "{Quantum Mechanics}: {Non-Relativistic Theory}",
    isbn = "978-0-7506-3539-4",
    publisher = "Butterworth-Heinemann",
    address = "Oxford",
    series = "Course of Theoretical Physics",
    volume = "v.3",
    year = "1991"
}

@book{Peskin:1995ev,
    author = "Peskin, Michael E. and Schroeder, Daniel V.",
    title = "{An Introduction to quantum field theory}",
    isbn = "978-0-201-50397-5",
    publisher = "Addison-Wesley",
    address = "Reading, USA",
    year = "1995"
}

@article{Maruyama:2021ghf,
    author = "Maruyama, Tomoyuki and Balantekin, A. Baha and Cheoung, Myung-Ki and Kajino, Toshitaka and Kusakabe, Motohiko and Mathewsh, Grant J.",
    title = "{A relativistic quantum approach to neutrino and antineutrino emission via the direct Urca process in strongly magnetized neutron-star matter}",
    eprint = "2103.01703",
    archivePrefix = "arXiv",
    primaryClass = "nucl-th",
    doi = "10.1016/j.physletb.2021.136813",
    journal = "Phys. Lett. B",
    volume = "824",
    pages = "136813",
    year = "2022"
}

@article{Duan:2005fc,
    author = "Duan, Huaiyu and Qian, Yong-Zhong",
    title = "{Rates of neutrino absorption on nucleons and the reverse processes in strong magnetic fields}",
    eprint = "astro-ph/0506033",
    archivePrefix = "arXiv",
    doi = "10.1103/PhysRevD.72.023005",
    journal = "Phys. Rev. D",
    volume = "72",
    pages = "023005",
    year = "2005"
}

@article{Mosta:2015ucs,
    author = {M\"osta, Philipp and Ott, Christian D. and Radice, David and Roberts, Luke F. and Schnetter, Erik and Haas, Roland},
    title = "{A large scale dynamo and magnetoturbulence in rapidly rotating core-collapse supernovae}",
    eprint = "1512.00838",
    archivePrefix = "arXiv",
    primaryClass = "astro-ph.HE",
    doi = "10.1038/nature15755",
    journal = "Nature",
    volume = "528",
    pages = "376",
    year = "2015"
}

@article{Raynaud:2020ist,
    author = {Raynaud, Rapha\"el and Guilet, J\'er\^ome and Janka, Hans-Thomas and Gastine, Thomas},
    title = "{Magnetar formation through a convective dynamo in protoneutron stars}",
    eprint = "2003.06662",
    archivePrefix = "arXiv",
    primaryClass = "astro-ph.HE",
    doi = "10.1126/sciadv.aay2732",
    journal = "Sci. Adv.",
    volume = "6",
    number = "11",
    pages = "eaay2732",
    year = "2020"
}

@article{Obergaulinger:2020cqq,
    author = "Obergaulinger, Martin and Aloy, Miguel-\'Angel",
    title = "{Magnetorotational core collapse of possible GRB progenitors. III. Three-dimensional models}",
    eprint = "2008.07205",
    archivePrefix = "arXiv",
    primaryClass = "astro-ph.HE",
    doi = "10.1093/mnras/stab295",
    journal = "Mon. Not. Roy. Astron. Soc.",
    volume = "503",
    number = "4",
    pages = "4942--4963",
    year = "2021"
}

@article{Mosta:2014jaa,
    author = {M\"osta, Philipp and Richers, Sherwood and Ott, Christian D. and Haas, Roland and Piro, Anthony L. and Boydstun, Kristen and Abdikamalov, Ernazar and Reisswig, Christian and Schnetter, Erik},
    title = "{Magnetorotational Core-Collapse Supernovae in Three Dimensions}",
    eprint = "1403.1230",
    archivePrefix = "arXiv",
    primaryClass = "astro-ph.HE",
    doi = "10.1088/2041-8205/785/2/L29",
    journal = "Astrophys. J. Lett.",
    volume = "785",
    pages = "L29",
    year = "2014"
}

@article{Famiano:2020fbq,
    author = "Famiano, Michael and Balantekin, A. Baha and Kajino, Toshitaka and Kusakabe, Motohiko and Mori, Kanji and Luo, Yudong",
    title = "{Nuclear Reaction Screening, Weak Interactions, and r-Process Nucleosynthesis in High Magnetic Fields}",
    eprint = "2006.14148",
    archivePrefix = "arXiv",
    primaryClass = "astro-ph.HE",
    doi = "10.3847/1538-4357/aba04d",
    journal = "Astrophys. J.",
    volume = "898",
    number = "2",
    pages = "163",
    year = "2020"
}

@article{OConnor:2009iuz,
    author = "O'Connor, Evan and Ott, Christian D.",
    editor = "Ott, Christian D. and Pethick, C. J. and Rezzolla, Luciano",
    title = "{A New Open-Source Code for Spherically-Symmetric Stellar Collapse to Neutron Stars and Black Holes}",
    eprint = "0912.2393",
    archivePrefix = "arXiv",
    primaryClass = "astro-ph.HE",
    doi = "10.1088/0264-9381/27/11/114103",
    journal = "Class. Quant. Grav.",
    volume = "27",
    pages = "114103",
    year = "2010"
}

@article{OConnor:2014sgn,
    author = "O'Connor, Evan",
    title = "{An Open-Source Neutrino Radiation Hydrodynamics Code for Core-Collapse Supernovae}",
    eprint = "1411.7058",
    archivePrefix = "arXiv",
    primaryClass = "astro-ph.HE",
    doi = "10.1088/0067-0049/219/2/24",
    journal = "Astrophys. J. Suppl.",
    volume = "219",
    number = "2",
    pages = "24",
    year = "2015"
}

@article{Shibata:2011kx,
    author = "Shibata, Masaru and Kiuchi, Kenta and Sekiguchi, Yu-ichiro and Suwa, Yudai",
    title = "{Truncated Moment Formalism for Radiation Hydrodynamics in Numerical Relativity}",
    eprint = "1104.3937",
    archivePrefix = "arXiv",
    primaryClass = "astro-ph.HE",
    doi = "10.1143/PTP.125.1255",
    journal = "Prog. Theor. Phys.",
    volume = "125",
    pages = "1255--1287",
    year = "2011"
}

@article{Janka:2006fh,
    author = "Janka, Hans-Thomas and Langanke, K. and Marek, A. and Martinez-Pinedo, G. and Mueller, B.",
    title = "{Theory of Core-Collapse Supernovae}",
    eprint = "astro-ph/0612072",
    archivePrefix = "arXiv",
    doi = "10.1016/j.physrep.2007.02.002",
    journal = "Phys. Rept.",
    volume = "442",
    pages = "38--74",
    year = "2007"
}

@article{Janka:2017vlw,
    author = "Janka, H. -Th.",
    title = "{Neutrino Emission from Supernovae}",
    journal = {},
    eprint = "1702.08713",
    archivePrefix = "arXiv",
    primaryClass = "astro-ph.HE",
     url = {https://doi.org/10.1007/978-3-319-21846-5_4},
    month = "2",
    year = "2017"
}

@article{OConnor:2012bsj,
    author = "O'Connor, Evan and Ott, Christian D.",
    title = "{The Progenitor Dependence of the Preexplosion Neutrino Emission in Core-Collapse Supernovae}",
    eprint = "1207.1100",
    archivePrefix = "arXiv",
    primaryClass = "astro-ph.HE",
    doi = "10.1088/0004-637X/762/2/126",
    journal = "Astrophys. J.",
    volume = "762",
    pages = "126",
    year = "2013"
}

@article{Duan:2004nc,
    author = "Duan, Huaiyu and Qian, Yong-Zhong",
    title = "{Neutrino processes in strong magnetic fields and implications for supernova dynamics}",
    eprint = "astro-ph/0401634",
    archivePrefix = "arXiv",
    doi = "10.1103/PhysRevD.69.123004",
    journal = "Phys. Rev. D",
    volume = "69",
    pages = "123004",
    year = "2004"
}

@article{Bethe:1990mw,
    author = "Bethe, H. A.",
    title = "{Supernova mechanisms}",
    doi = "10.1103/RevModPhys.62.801",
    journal = "Rev. Mod. Phys.",
    volume = "62",
    pages = "801--866",
    year = "1990"
}

@article{Lattimer:1991nc,
    author = "Lattimer, James M. and Swesty, F. Douglas",
    title = "{A Generalized equation of state for hot, dense matter}",
    doi = "10.1016/0375-9474(91)90452-C",
    journal = "Nucl. Phys. A",
    volume = "535",
    pages = "331--376",
    year = "1991"
}

@misc{Heger_pri, author = {Alex Heger}, year = {Private Communication}}

@ARTICLE{sumiyoshi06,
       author = {{Sumiyoshi}, K. and {Yamada}, S. and {Suzuki}, H. and {Chiba}, S.},
        title = "{Neutrino Signals from the Formation of a Black Hole: A Probe of the Equation of State of Dense Matter}",
      journal = {\prl},
         year = 2006,
        month = sep,
       volume = {97},
       number = {9},
          eid = {091101},
        pages = {091101},
          doi = {10.1103/PhysRevLett.97.091101},
archivePrefix = {arXiv},
       eprint = {astro-ph/0608509},
 primaryClass = {astro-ph},
       adsurl = {https://ui.adsabs.harvard.edu/abs/2006PhRvL..97i1101S}
}

@ARTICLE{kitaura06,
       author = {{Kitaura}, F.~S. and {Janka}, H. -Th. and {Hillebrandt}, W.},
        title = "{Explosions of O-Ne-Mg cores, the Crab supernova, and subluminous type II-P supernovae}",
      journal = {Astron. Astrophys.},
         year = 2006,
        month = apr,
       volume = {450},
       number = {1},
        pages = {345-350},
          doi = {10.1051/0004-6361:20054703},
archivePrefix = {arXiv},
       eprint = {astro-ph/0512065},
 primaryClass = {astro-ph},
       adsurl = {https://ui.adsabs.harvard.edu/abs/2006A&A...450..345K}
}

@ARTICLE{radice17,
       author = {{Radice}, David and {Burrows}, Adam and {Vartanyan}, David and {Skinner}, M. Aaron and {Dolence}, Joshua C.},
        title = "{Electron-capture and Low-mass Iron-core-collapse Supernovae: New Neutrino-radiation-hydrodynamics Simulations}",
      journal = {\apj},
         year = 2017,
        month = nov,
       volume = {850},
       number = {1},
          eid = {43},
        pages = {43},
          doi = {10.3847/1538-4357/aa92c5},
archivePrefix = {arXiv},
       eprint = {1702.03927},
 primaryClass = {astro-ph.HE},
       adsurl = {https://ui.adsabs.harvard.edu/abs/2017ApJ...850...43R}
}

@ARTICLE{powell23,
       author = {{Powell}, Jade and {M{\"u}ller}, Bernhard and {Aguilera-Dena}, David R. and {Langer}, Norbert},
        title = "{Three dimensional magnetorotational core-collapse supernova explosions of a 39 solar mass progenitor star}",
      journal = {Mon. Not. Roy. Astron. Soc.},
         year = 2023,
        month = jul,
       volume = {522},
       number = {4},
        pages = {6070-6086},
          doi = {10.1093/mnras/stad1292},
archivePrefix = {arXiv},
       eprint = {2212.00200},
 primaryClass = {astro-ph.HE},
       adsurl = {https://ui.adsabs.harvard.edu/abs/2023MNRAS.522.6070P}
}

@ARTICLE{fischer10,
       author = {{Fischer}, T. and {Whitehouse}, S.~C. and {Mezzacappa}, A. and {Thielemann}, F. -K. and {Liebend{\"o}rfer}, M.},
        title = "{Protoneutron star evolution and the neutrino-driven wind in general relativistic neutrino radiation hydrodynamics simulations}",
      journal = {Astron. Astrophys.},
         year = 2010,
        month = jul,
       volume = {517},
          eid = {A80},
        pages = {A80},
          doi = {10.1051/0004-6361/200913106},
archivePrefix = {arXiv},
       eprint = {0908.1871},
 primaryClass = {astro-ph.HE},
       adsurl = {https://ui.adsabs.harvard.edu/abs/2010A&A...517A..80F}
}

@ARTICLE{oconnor11,
       author = {{O'Connor}, Evan and {Ott}, Christian D.},
        title = "{Black Hole Formation in Failing Core-Collapse Supernovae}",
      journal = {\apj},
         year = 2011,
        month = apr,
       volume = {730},
       number = {2},
          eid = {70},
        pages = {70},
          doi = {10.1088/0004-637X/730/2/70},
archivePrefix = {arXiv},
       eprint = {1010.5550},
 primaryClass = {astro-ph.HE},
       adsurl = {https://ui.adsabs.harvard.edu/abs/2011ApJ...730...70O}
}

@ARTICLE{oertel17,
       author = {{Oertel}, M. and {Hempel}, M. and {Kl{\"a}hn}, T. and {Typel}, S.},
        title = "{Equations of state for supernovae and compact stars}",
      journal = {Reviews of Modern Physics},
         year = 2017,
        month = jan,
       volume = {89},
       number = {1},
          eid = {015007},
        pages = {015007},
          doi = {10.1103/RevModPhys.89.015007},
archivePrefix = {arXiv},
       eprint = {1610.03361},
 primaryClass = {astro-ph.HE},
       adsurl = {https://ui.adsabs.harvard.edu/abs/2017RvMP...89a5007O}
}

@ARTICLE{2021ApJ...906..128K,
       author = {{Kuroda}, Takami},
        title = "{Impact of a Magnetic Field on Neutrino-Matter Interactions in Core-collapse Supernovae}",
      journal = {\apj},
         year = 2021,
        month = jan,
       volume = {906},
       number = {2},
          eid = {128},
        pages = {128},
          doi = {10.3847/1538-4357/abce61},
archivePrefix = {arXiv},
       eprint = {2009.07733},
 primaryClass = {astro-ph.HE},
       adsurl = {https://ui.adsabs.harvard.edu/abs/2021ApJ...906..128K}
}
\end{document}